\documentstyle[psfig]{mn}

\newif\ifAMStwofonts

\newcommand{\FeH}{\mbox{[Fe/H]}\,}

\ifoldfss

  \ifCUPmtlplainloaded \else

    \NewTextAlphabet{textbfit} {cmbxti10} {}

    \NewTextAlphabet{textbfss} {cmssbx10} {}

    \NewMathAlphabet{mathbfit} {cmbxti10} {} 

    \NewMathAlphabet{mathbfss} {cmssbx10} {} 

  \fi

  \ifAMStwofonts

    \ifCUPmtlplainloaded \else

      \NewSymbolFont{upmath} {eurm10}

      \NewSymbolFont{AMSa} {msam10}

      \NewMathSymbol{\upi}     {0}{upmath}{19}

      \NewMathSymbol{\umu}     {0}{upmath}{16}

      \NewMathSymbol{\upartial}{0}{upmath}{40}

      \NewMathSymbol{\leqslant}{3}{AMSa}{36}

      \NewMathSymbol{\geqslant}{3}{AMSa}{3E}

       \let\le=\leqslant

    \fi

  \fi

\fi 

\ifnfssone

  \newmathalphabet{\mathit}

  \addtoversion{normal}{\mathit}{cmr}{m}{it}

  \addtoversion{bold}{\mathit}{cmr}{bx}{it}

  \newmathalphabet{\mathbfss} 

  \addtoversion{normal}{\mathbfss}{cmss}{bx}{n}

  \addtoversion{bold}{\mathbfss}{cmss}{bx}{n}

  \ifAMStwofonts

    \ifCUPmtlplainloaded \else

      \UseAMStwoboldmath

      \makeatletter

      \new@mathgroup\upmath@group

      \define@mathgroup\mv@normal\upmath@group{eur}{m}{n}

      \define@mathgroup\mv@bold\upmath@group{eur}{b}{n}

      \edef\UPM{\hexnumber\upmath@group}

      \new@mathgroup\amsa@group

      \define@mathgroup\mv@normal\amsa@group{msa}{m}{n}

      \define@mathgroup\mv@bold\amsa@group{msa}{m}{n}

      \edef\AMSa{\hexnumber\amsa@group}

      \makeatother

      \mathchardef\upi="0\UPM19

      \mathchardef\umu="0\UPM16

      \mathchardef\upartial="0\UPM40

      \mathchardef\leqslant="3\AMSa36

      \mathchardef\geqslant="3\AMSa3E

       \let\le=\leqslant

    \fi

  \fi

\fi 

\ifnfsstwo

  \DeclareMathAlphabet{\mathbfit}{OT1}{cmr}{bx}{it}

  \SetMathAlphabet\mathbfit{bold}{OT1}{cmr}{bx}{it}

  \DeclareMathAlphabet{\mathbfss}{OT1}{cmss}{bx}{n}

  \SetMathAlphabet\mathbfss{bold}{OT1}{cmss}{bx}{n}

  \ifAMStwofonts

    \ifCUPmtlplainloaded \else

      \DeclareSymbolFont{UPM}{U}{eur}{m}{n}

      \SetSymbolFont{UPM}{bold}{U}{eur}{b}{n}      \DeclareSymbolFont{AMSa}{U}{msa}{m}{n}

      \DeclareMathSymbol{\upi}{0}{UPM}{"19}

      \DeclareMathSymbol{\umu}{0}{UPM}{"16}

      \DeclareMathSymbol{\upartial}{0}{UPM}{"40}

      \DeclareMathSymbol{\leqslant}{3}{AMSa}{"36}

      \DeclareMathSymbol{\geqslant}{3}{AMSa}{"3E}

       \let\le=\leqslant

    \fi

  \fi

\fi 

\ifCUPmtlplainloaded \else

  \ifAMStwofonts \else 

    \def\upi{\pi}

    \def\umu{\mu}

    \def\upartial{\partial}

  \fi

\fi

\title{A pulsational approach to near infrared and
visual magnitudes of RR Lyrae stars}

\author[G. Bono et al.]
 {G. Bono$^1$, F. Caputo$^1$, V. Castellani$^{1,2}$, M. Marconi$^3$,
J. Storm$^4$, S. Degl'Innocenti$^5$\\
   $^1$ INAF-Osservatorio Astronomico di Roma, via di Frascati 33, 00040 Monte
        Porzio Catone, Italy~~bono/caputo/vittorio@mporzio.astro.it\\
   $^2$ INFN - Sezione di Ferrara, via Paradiso 12, 44100 Ferrara, Italy \\
   $^3$ INAF-Osservatorio Astronomico di Capodimonte, via Moiariello 16, 80131
        Napoli, Italy~~marcella@na.astro.it\\
   $^4$ Astrophysikalisches Institut Potsdam, An der Sternwarte 16,
        14482 Potsdam, Germany~~jstorm@aip.de\\
   $^5$ Dipartimento di Fisica, Universit\`a di Pisa,
via Buonarroti 2, 56127, Pisa,
   Italy~~scilla@df.unipi.it}

\date{}

\begin{document}
\maketitle

\label{firstpage}

\begin{abstract}
In this paper we present an improved theoretical scenario
concerning near infrared and visual magnitudes of RR Lyrae
variables, as based on up-to-date pulsating models. New relations
connecting $V$ and $K$ absolute magnitudes with periods, mass,
luminosity and metal content are discussed separately for
fundamental and first overtone pulsators. We also show that the
$V-K$ colors are predicted to supply tight constraints on the
pulsator intrinsic luminosity. On this basis, we revisit the case
of the prototype variable RR Lyr, showing that the parallax
inferred by this new pulsational approach appears in close
agreement with HST absolute parallax. Moreover, available $K$ and
$V$ measurements for field and cluster RR Lyrae variables with
known reddening and metal content are used to derive a relation
connecting the $K$ absolute magnitude to period and metallicity
($M_K$-[Fe/H]-log$P$) as well as a new calibration of the
$M_V$-[Fe/H] relation. The comparison between theoretical
prescriptions and observations suggests that RR Lyrae stars in the
field and in Galactic Globular Clusters (GGCs) should have quite
similar evolutionary histories. The comparison between theory and
observations also discloses a general agreement that supports the
reliability of current pulsational scenario. On the contrary,
current empirical absolute magnitudes based on the Baade-Wesselink
(BW) method suggest relations with a zero-point that is fainter
than predicted by pulsation models, together with a milder
metallicity dependence. However, preliminary results based on a
new calibration of the BW method provided by Cacciari et al.
(2000) for RR Cet and SW And appear in a much better agreement
with the pulsational predictions.
\end{abstract}


\section{Introduction}
Cosmic distances are fundamental cornerstones in modern astronomy,
and reliable distance indicators can supply tight constraints on
several cosmological parameters (see e.g. Goobar et al. 2000;
Ferrarese et al. 2000). Accordingly, the number of theoretical and
empirical investigations looking for new standard candles and/or
aimed at improving the classical ones is countless in the recent
literature (see e.g. Salaris \& Cassisi 1997; Bono et al. 1999;
Feast 1999; Walker 1999; Caputo et al. 2000; Carretta et al. 2000;
Castellani et al. 2000; Gieren, Fouqu\'e, \& Storm 2000; Layden
2002). However, the ongoing problems affecting distance
determinations are soundly revealed by current uncertainties on
the true distance modulus $\mu_{0}$ of the nearest galaxy, the
Large Magellanic Cloud (LMC), which is often taken as a reference
for extragalactic distance determinations. In fact, current
estimates range from $\mu_{0,LMC}\approx18.37\pm0.23$ mag (RR
Lyrae, Luri et al. 1998) to $\approx$ 18.7 mag (classical
Cepheids, Feast \& Catchpole 1997; Bono et al. 1999; Caputo et al.
2002).

In this paper we discuss RR Lyrae stars, the radial pulsators
connected with low-mass He-burning stars in the Horizontal Branch
(HB) evolutionary phase. The use of these variables as distance
indicators appears still hampered by significant uncertainties,
and indeed the relation between absolute visual magnitude and iron
content, $M_V$-[Fe/H], calibrated according to different empirical
or theoretical constraints, supplies absolute magnitudes that for
a given metallicity cover $\approx$ 0.3 mag. Oddly enough, one
also finds that the empirical approach based on the BW method,
gives different results when different bands (optical vs near
infrared) or different approaches (surface brightness vs canonical
BW) are used.

In this context, the endeavor to link RR Lyrae pulsation
properties, i.e. periods and/or amplitudes of the light curve, to
the absolute magnitudes of these stars (see Sandage 1993; Caputo
1997; Caputo et al. 2000; Caputo, Degl'Innocenti \& Marconi 2001)
appear of great relevance, because periods (and amplitudes) are
firm and safe observational parameters, independent of distance
and reddening. Empirical investigations (Longmore, Fernley, \&
Jameson 1986; Longmore et al. 1990) have already suggested that in
the $K$ band the RR Lyrae stars in GGCs follow  a rather
well-defined Period-Luminosity ($PL_K$) relation whose slope is,
within observational errors, almost constant when moving from
metal-poor to metal-rich clusters. This scenario has been recently
supported by the theoretical investigation by Bono et al. (2001,
Paper I). In that paper, on the basis of nonlinear, convective RR
Lyrae models, we found that the  $K$-magnitude of fundamental (F)
and fundamentalized first-overtone (FO) pulsators can be fitted by
a common Period-Luminosity-Metallicity ($PLZ_K$) relation which
appears to be in good agreement with observations.

In this paper we present improved predictions concerning both F
and FO pulsators. Section 2 presents new theoretical relations
concerning $V$ and $K$ absolute magnitudes. On this basis, in the
same section we discuss improved $PLZ_K$ relations, showing that
$V-K$ colors might allow us to constrain the intrinsic bolometric
luminosity of the pulsators. Section 3 deals with the prototype
variable RR Lyr, used as fundamental calibrator of the RR Lyrae
distance scale, while observational data  for field and cluster RR
Lyrae stars are discussed in Section 4. In Section 5 we present a
new relation connecting the $K$ absolute magnitude to period and
metallicity ($M_K$-[Fe/H]-log$P$) as well as a refined
$M_V$-[Fe/H] relation. A few final remarks concerning both
theoretical models and observations are outlined in Section 6

\section{New $PLZ_K$ and $PLC_{VK}$ relations for F and FO pulsators}

To properly implement the pulsational framework presented in Paper
I, we computed additional pulsating models adopting a finer grid
of effective temperatures. We ended up with 132 F and 91 FO
nonlinear models that cover a wide metallicity range $0.0001 \le Z
\le 0.02$. For each metal abundance the helium content was fixed
according to a primordial helium abundance of $Y_p$=0.24 and a
helium-to-metal enrichment ratio $\Delta Y/\Delta Z \approx $ 2.5.

At fixed chemical composition, we explored three different
luminosity levels that  cover current uncertainties on the ZAHB
luminosity as well as post-ZAHB evolutionary effects (see column 3
in Table 1). As for as the pulsator mass is concerned, we adopted
for each given metallicity a "reference mass" $M_r$, roughly
following the correlation between metal content and mass of ZAHB
models populating the RR Lyrae instability strip, i.e. $3.77 \le
\log T_e \le 3.86$. These $M_r$ values are listed in Table 1
(column 4), together with the evolutionary mass (with an average
uncertainty of $\sim$ 2\%) of ZAHB models at log$T_e$=3.85 and
3.80, as inferred according to evolutionary computations available
in the literature. In order to investigate  the dependence of the
predicted relations on the pulsator mass, additional models were
finally constructed at selected metal contents ($Z=0.0001$ and
$Z=0.001$) but with slightly different reference mass values.

\begin{table}
\centering
\caption[]{Input parameters for the grid of RR Lyrae models.
Mass and luminosity are in solar units.\label{tab1}}
\begin{tabular}{llcccc}
 $Z$  & $Y$ & log $L$  & $M_r$ & $M(3.85)$ & $M(3.80)$ \\
      &     &       &          &  $\pm$2\% &  $\pm$2\%\\
0.0001 & 0.24  & 1.61, 1.72, 1.81 & 0.75 & 0.796 & 0.852  \\
0.0004 & 0.24  & 1.61, 1.72, 1.81 & 0.70 & 0.699 & 0.721   \\
0.001  & 0.24  & 1.51, 1.61, 1.72 & 0.65 & 0.648 & 0.666 \\
0.006  & 0.255 & 1.55, 1.65, 1.75 & 0.58 & 0.585 & 0.589 \\
0.01   & 0.255 & 1.51, 1.57, 1.65 & 0.58 & 0.575 & 0.578 \\
0.02   & 0.28  & 1.41, 1.51, 1.61 & 0.53 & 0.545 & 0.546 \\
\end{tabular}
\end{table}

The entire set of fundamental and first overtone nonlinear models were
integrated in time till they approach the limit cycle stability. These
calculations provided periods, pulsation amplitudes, and bolometric 
light curves. The light curves were transformed into the observational 
plane by adopting bolometric corrections and
color-temperature relations provided by Castelli, Gratton \&
Kurucz (1997a,b). Tables 2 and 3 list the predicted periods (in
days) and mean (magnitude-weighted) $V$ and $K$ magnitudes for the
new F and FO pulsation models, respectively. These new data
implement similar predictions already presented in Tables 2 and 3
of Paper I.

\begin{table}
\caption[]{Periods and absolute mean (magnitude-weighted)
magnitudes for the new fundamental models.\label{tab2}}

\begin{tabular}{ccccrc}

\hline

 $Z$ & log $L/L_{\odot}$ & $T_e$ & $P$ & $M_K$ & $M_V$ \\

     &                   & (K)   &  (d)& (mag) & (mag)  \\

\hline

0.0001 &  1.61 &  6800 & 0.3876 &-0.172 &0.848  \\

0.0001 &  1.61 &  6700 &0.4077 &-0.226 &0.845  \\

0.0001 &  1.61 &  6600 &0.4291 &-0.282 &0.844  \\

0.0001 &  1.61 &  6500 &0.4518 &-0.328 &0.834  \\

0.0001 &  1.61 &  6400 &0.4764 &-0.378 &0.833  \\

0.0001 &  1.61 &  6200 &0.5299 &-0.470 &0.843  \\

0.0001 &  1.61 &  6100 &0.5594 &-0.507 &0.848  \\

0.0001 &  1.72 &  6800  &  0.4790    &  -0.448  &  0.597 \\

0.0001 &  1.72 &  6600   &  0.5292    &  -0.554  &  0.572             \\

0.0001 &  1.72 &  6400   &  0.5882    &  -0.658  &  0.562           \\

0.0001 &  1.72 &  6200   &   0.6561   &  -0.752  &  0.566           \\

0.0001 &  1.81 &  6600   &   0.6304   &  -0.778  &  0.357          \\

0.0001 &  1.81 &  6400   &   0.7005   &  -0.880  &  0.338           \\

0.0001 &  1.81 &  6300   &   0.7399   &  -0.930  &  0.335          \\

0.0004 &  1.81 &  6600   &   0.6632   &  -0.778  &  0.330        \\

0.0004 &  1.81 &  6500   &   0.7014   &  -0.830 &  0.320          \\

0.0004 &  1.81 &  6400   &   0.7400  &   -0.881  &  0.315           \\

0.006  &  1.75 &  6900    &   0.5988    &   -0.494    &  0.428 \\

0.006  &  1.75 &  6800    &   0.6290   &    -0.546    &  0.426 \\

0.006  &  1.75 &  6700   &    0.6614   &    -0.598    &  0.416 \\

0.006  &  1.75 &  6600   &    0.6972   &    -0.650    &  0.408 \\

0.006  &  1.75 &  6500   &    0.7349    &   -0.702    & 0.402 \\

0.006  &  1.75 &  6400    &   0.7755    &   -0.753    &  0.399 \\

0.006  &  1.75 &  6300   &    0.8184    &   -0.803    &  0.400 \\

0.006  &  1.75 &  6200   &    0.8626    &   -0.850   &  0.403 \\

0.006  &  1.75 &  6000    &   0.9632    &   -0.962     & 0.404 \\

0.006  &  1.75 &  5900    &   1.0184     &  -0.988    &  0.423 \\

0.006  &  1.75 &  5800    &   1.0794     &  -1.029     & 0.430  \\

0.01   &  1.65 &   6800    &   0.5162    &     -0.300    &    0.697 \\

0.01   &  1.65 &   6600    &   0.5708    &     -0.403    &    0.682 \\

0.01   &  1.65 &   6500    &   0.6019   &      -0.455    &    0.672 \\

0.01   &  1.65 &   6400     &  0.6322     &    -0.509    &    0.668 \\

0.01   &  1.65 &   6300     &  0.6669     &    -0.558     &   0.665 \\

0.01   &  1.65 &  6200     &  0.7052     &       -0.607   &     0.664 \\

0.01   &  1.65 &  6100     &  0.7441      &     -0.658     &   0.668 \\

0.01   &  1.65 & 5900     &  0.8345      &    -0.762       &   0.681 \\

0.01   &  1.65 & 5700     &  0.9331      &    -0.839    &    0.695 \\

\hline

\hline

\end{tabular}

\end{table}

\begin{table}

\caption[]{Periods and absolute mean (magnitude-weighted) magnitudes

for the new first overtone models.\label{tab3}}

\begin{tabular}{ccccrc}

\hline

 $Z$ &$\log L/L_{\odot}$ & $T_e$ & $P$ & $M_K$ & $M_V$ \\

     &                   & (K)   & (d) & (mag) & (mag)  \\

\hline

0.0001 &  1.61  & 7100  &        0.2546  &      -0.031      &   0.817 \\

0.0001 &  1.61  & 7000  &       0.2657   &      -0.081      &   0.817 \\

0.0001 &  1.61  & 6900  &        0.2784  &       -0.131     &    0.812 \\

0.0001 &  1.61  & 6800  &       0.2915   &      -0.180      &   0.808 \\

0.0001 &  1.61  & 6600  &       0.2935   &      -0.276      &   0.806  \\

0.006  &  1.75  & 7000  &     0.4132     &      -0.453     &      0.325 \\

0.006  &  1.75  & 6900  &     0.4349     &      -0.502     &      0.350 \\

0.006  &  1.75  & 6800  &     0.4563     &      -0.552      &     0.363 \\

0.006  &  1.75  & 6700  &     0.4806     &      -0.603     &      0.370 \\

0.006  &  1.75  & 6600  &     0.5070     &      -0.656     &      0.374 \\

0.006  &  1.75  & 6500  &    0.5343      &     -0.709      &     0.378 \\

0.006  &  1.75  & 6400  &      0.5615    &       -0.759    &       0.383 \\

0.006  &  1.75  & 6300  &       0.5904    &       -0.806      &     0.389 \\

0.006  &  1.75  & 6200  &     0.6222      &     -0.849        &   0.394   \\

0.01   &  1.51  & 6800  &       0.2871      &      0.045      &     0.965 \\

0.01   &  1.57  & 6900  &      0.3070     &     -0.055     &     0.800 \\

0.01   &  1.57  & 6800  &     0.3228      &    -0.104      &    0.807   \\

0.01   &  1.65  & 7000  &        0.3394     &      -0.204      &     0.574  \\

0.01   &  1.65  & 6900  &       0.3562     &      -0.253       &    0.594  \\

0.01   &  1.65  & 6800  &      0.3745    &       -0.302       &    0.603  \\

0.01   &  1.65  & 6700  &       0.3931     &      -0.355       &    0.611  \\

0.01   &  1.65  & 6600  &      0.4135     &      -0.402       &    0.615    \\

\hline

\hline

\end{tabular}

\end{table}

Data listed in Table 2 and Table 3 disclose that, for any given
mass, luminosity and metal content, the visual magnitude $M_V$ of
the pulsators is quite independent of the effective temperature,
whereas $M_K$ shows a strong dependence on this parameter (see
also Paper I). This effect is the consequence of the fact that in
the range of RR Lyrae effective temperatures the visual bolometric
correction is roughly constant, whereas in the $K$-band it is
strongly dependent on the effective temperature. It follows that
the $K$-band magnitudes become significantly brighter when moving
toward cooler effective temperatures. The pulsation equation
$P=f(L,M,T_e)$ connects, for each given chemical composition,
periods to luminosity, mass and effective temperature. Therefore,
the dependence of the absolute magnitude on $T_e$ (through the
$BC$) can be replaced with a dependence on period. Accordingly,
when moving from the blue to the red edge of the instability strip
the period increases at roughly constant absolute visual magnitude
($\delta M_V/\delta$log$P \sim -$0.14), whereas $M_K$ shows a much
steeper correlation ($\delta M_K/\delta$log$P \sim -$2.1.)

On the other hand, when varying the luminosity for any given value
of mass and effective temperature, one has again a relation
between absolute magnitude and period. The key feature of the
$K$-band is that the two derivatives ($\delta
M_K/\delta$log$P$)$_{M,L)}$ and ($\delta
M_K/\delta$log$P$)$_{M,T_e)}$ are quite similar. In other words,
one finds a sort of $degeneracy$ for which the near-infrared
period-magnitude ($PL_K$) relation is only marginally dependent on
the bolometric luminosity of pulsators (see Fig. 1 in Paper I). 

As for the dependence of the $PL_K$ relation on the pulsator mass and 
metal content, current calculations confirm the results presented in 
Paper I. However, at variance with the approach adopted in that paper, 
where F and fundamentalised FO periods
(i.e., $\log P_F=\log P_{FO} +0.127$) were taken simultaneously
into account, we found that the intrinsic accuracy can be improved
by independently treating the two pulsation modes.

By accounting also for the mass dependence, a linear regression 
through the models gives:

$$M_K^F= 0.511-2.102\log P+0.095\log Z -0.734\log L- 1.753\log M/M_r\eqno(1)$$
$$M_K^{FO}=-0.029-2.265\log P+0.087\log Z-0.635\log L-1.633\log M/M_r\eqno(2)$$

\noindent where mass and luminosity are in solar units and the rms
scatter is $\sigma$=0.016~mag. As already discussed in Paper I,
the predicted slope at constant luminosity and metallicity appears
in good agreement with empirical data for GGC variables, as given,
e.g., by the RR Lyrae stars in M3 collected by Longmore et al.
(1990).

The present enlarged set of pulsating models confirms that over
the explored range of metallicity and luminosity the $M_K -\log P$
correlation is only mildly dependent on the luminosity, a
variation of 0.1 dex implying a change of $\delta M_K\sim 0.07$
mag, as well as on the metallicity, a variation of 0.3 dex
implying a change of $\delta M_K\sim 0.03$ mag. The predicted
magnitudes are thus expected to be only mildly dependent on
uncertainties introduced by these theoretical or observational
parameters. On the other hand, the non negligible dependence on
the pulsator mass should only introduce a marginal uncertainty in
the error budget since we can use quite firm theoretical
constraints on such evolutionary parameter. As a fact,  HB models
available in the recent literature supply slightly discordant
luminosity values, but very similar predictions concerning the
mass.

As well known, fundamental ($RRab$) and first-overtone ($RRc$) RR
Lyrae populate the red and the blue side of the pulsation region,
respectively. Therefore one may assume for F and FO pulsators the
predicted stellar mass listed in Table 1 for ZAHB models at $\log
T_e=3.80$ and 3.85, respectively. The total uncertainty introduced
by this assumption is of the order of 3.5\%  if we simultaneously
account for the predicted spread in mass between F and FO
pulsators and for the occurrence of post-ZAHB evolutionary
effects.

By inserting these mass values into the previous relations, one
eventually derives the following near-infrared
Period-Luminosity-Metallicity ($PLZ_K$) relations:

$$M_K^F= 0.565-2.101\log P+0.125\log Z-0.734\log L\eqno(3)$$
$$M_K^{FO}=-0.016-2.265\log P+0.096\log Z-0.635\log L\eqno(4)$$

\noindent with a rms scatter of $\sigma_K=0.031$ mag and 0.025
mag, respectively, including in quadrature an uncertainty of
$\sim$ 0.03 mag, as given by the adopted spread in mass (3.5\%) at
fixed metal abundance.

By following  a similar procedure, one finds that also the $T_e$
dependence of the pulsator color can be replaced by a period
dependence. In particular,  since the $V$ magnitudes are strongly
correlated with $L$, whereas the $K$ magnitudes do not, we expect
that the $V-K$ color is strongly dependent on luminosity, thus
providing useful constraints to this unknown parameter.

As a fact, using the whole set of pulsating models, we derive:

$$(M_V-M_K)^F=4.014+1.986\log P-0.134\log Z-1.662\log L+1.656\log M/M_r \eqno(5)$$

$$(M_V-M_K)^{FO}=5.195+2.518\log P-0.159\log Z-2.158\log L+1.815\log M/M_r \eqno(6)$$

\noindent with a rms scatter of 0.027 mag and 0.023 mag,
respectively (see Fig. 1).

\begin{figure}
\psfig{figure=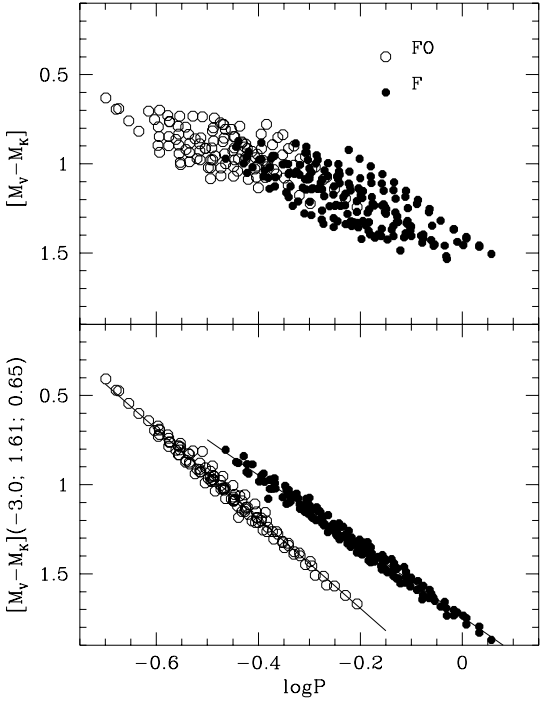,height=10cm}
\caption{Top panel: predicted $M_V-M_K$ colors of F  (filled circles) and
FO pulsators (open circles)
as a function of period (data in Table 2 and Table 3).
Bottom panel: same as the
top panel but with individual colors scaled
to the same metallicity ($\log Z=-3$), luminosity ($\log L/L_\odot=1.61$),
and mass (0.65$M{_\odot}$). The solid lines display
the predicted relations (see equation (5) and (6) in the text).}
\end{figure}

By assuming the predicted masses at log$T_e$=3.80 and 3.85 for F
and FO pulsators, respectively, the previous relations yield the
Period-Luminosity-Color relations ($PLC_{VK}$) in the $V,K$ bands
for both F and FO pulsators, as given by:

$$(M_V-M_K)^F=3.963+1.986\log P-0.162\log Z-1.662\log L\eqno(7)$$
$$(M_V-M_K)^{FO}=5.180+2.518\log P-0.168\log Z-2.158\log L\eqno(8)$$

\noindent
with a total intrinsic dispersion of 0.037 mag and 0.031 mag, respectively,
which accounts for the adopted spread in mass (3.5\%) at fixed metallicity.

According to eqs. (7) and (8), one finds that accurate $V,K$
photometric measurements of F and FO pulsators with known period
and metal content can supply independent information on the
intrinsic luminosity of the variable. Moreover, since
$E(V-K)$=0.89$A_V$ (Cardelli et al. 1989), if the visual
extinction along the line of sight is known with an accuracy of
$\pm$ 0.03~mag, then the luminosity might be evaluated with a
(formal) accuracy of $\sim$ 0.03 dex.

\section{The case of RR Lyr itself}

As already discussed by Bono et al. (2002, Paper II), the
predicted $PLZ_K$ relation for F pulsators can be used to provide
for the prototype variable RR Lyr a  "pulsational" parallax that
appears in close agreement with the absolute value recently
measured by Benedict et al. (2002, hereafter B02) on the basis of
new astrometric data collected with FGS~3, the interferometer on
board of HST. In that paper, {\it we assumed} an intrinsic
luminosity of RR Lyr in the range log$L/L_{\odot}$=1.65-1.80 to
cover current uncertainties of HB models. On this basis, we
derived $M_K$=$-0.541\pm$0.062 mag and a  "pulsation" parallax
$\pi_{puls}$=3.858$\pm$0.131~mas, in close agreement with the HST
absolute value $\pi_{abs}$=3.82$\pm$0.20~mas.

According to the results presented in the previous section, one
may revisit the case of RR Lyr, avoiding any assumption on the RR
Lyr bolometric luminosity. Adopting log$P$=$-$0.2466 (Kazarovets,
Samus, \& Durlevich 2001), $K$=6.54 mag (Fernley, Skillen, \&
Burki 1993), [Fe/H]=$-$1.39$\pm0.15$ (Fernley et al. 1998a; Beers
et al. 2000) and $V$=7.784 mag (Hardie 1955), equation (7) gives
$\log L =1.642\pm0.024 + 0.535 A_V$. Thus, for each adopted
extinction correction, one $derives$ directly from the observed
$V-K$ color the luminosity value to be inserted into equation (3),
without using any evolutionary predictions. Once $M_K$ is
determined from equation (3), both the intrinsic distance modulus
($\mu_0=K-0.11A_V-M_K$) and the absolute visual magnitude
($M_V=V-A_V-\mu_0$) can be easily determined.

Data listed in Table 4 show that the pulsation parallax presents a
mild dependence on the adopted extinction correction. In particular,
the top panel in Fig. 2 shows that when moving from $A_V=0.01$ to
0.22 mag, the pulsation parallax (filled circles) decreases from
$3.89\pm0.07$ to $3.78\pm0.07$ mas. However, current pulsational
estimates remain well within the errors of the HST absolute
parallax (dashed lines), and present a better formal accuracy
($\sigma_\pi$/$\pi\sim$ 3\% versus $\sim$ 5\%). According to this
evidence the pulsation parallaxes resemble direct astrometric
measurements that are independent of interstellar absorption.

\begin{figure}
\psfig{figure=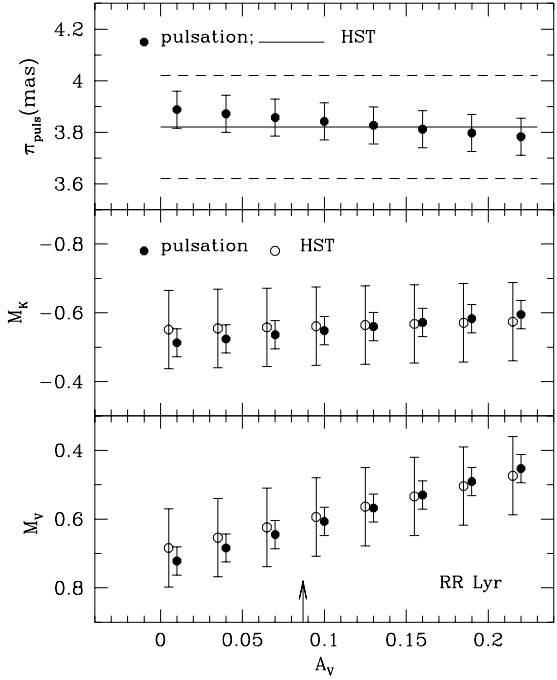,height=10cm}
\caption{Top panel: predicted
pulsation parallax for RR Lyr (filled circles) as a function of
the visual extinction. Solid and dashed lines show the HST direct
measurement and the relative uncertainty (B02), respectively. .
Middle panel: comparison between absolute $K$-magnitudes of RR Lyr
based on HST trigonometric parallax (open circles) and on the
pulsation approach (filled circles), as a function of the visual
extinction. For clarity sake, HST data are shifted by $-$0.005
along the x-axis. Bottom panel: the same, but for the absolute
visual magnitude. The arrow marks the weighted mean of current
extinction corrections}
\end{figure}

Concerning the RR Lyr predicted absolute magnitude, the middle and
the bottom panel in Fig. 2 show that when moving from $A_V$=0.01
to 0.22 mag $M_K$ changes from $-0.513\pm0.041$ mag to
$-0.595\pm0.041$ mag, while $M_V$ from $0.722\pm0.041$ to
$0.453\pm0.041$ mag, in close agreement with HST results (open
circles), but with a better formal accuracy ($\pm0.041$ versus
$\pm0.114$ mag). In other words, despite the substantial
improvement in the accuracy of the RR Lyr trigonometric parallax
provided by HST, when compared with previous measurements (see
B02), data plotted in Fig. 2 show that a sound empirical
determination of its absolute magnitude is still hampered by the
intrinsic uncertainty on the HST measurement, {\it even if the
interstellar extinction to RR Lyr would be firmly known}.

\begin{table}
\centering
\caption[]{The case of RR Lyr itself. For each adopted
extinction correction $A_V$, column 2 gives the luminosity (in solar
units) derived using the equation (7). This luminosity is used in equation
(3) to evaluate $M_K (3)$, and in turn to estimate together with $K=6.54$
mag the intrinsic distance modulus, $\mu_0$, and the "pulsational" parallax
$\pi_{puls}$. The last two columns give the absolute visual magnitude,
based on $V=7.784$ mag as well as on the pulsational and on the HST absolute
parallax ($\pi_{abs}$=3.82$\pm$0.20 mas), respectively.\label{tab4}}
\begin{tabular}{lcccccc}
\hline
$A_V$ & log$L(7)$ & $M_K(3)$ & $\mu_0$ &$\pi_{puls}$ &$M_V^{\pi_{puls}}$ & $M_V^{HST}$ \\

  (mag) & &  (mag) & (mag) &(mas) & (mag) & (mag) \\
  & $\pm$0.024 & $\pm$ 0.041 & $\pm$ 0.041 & $\pm$ 0.072& $\pm$ 0.041 & $\pm$ 0.114 \\

\hline
0.01 & 1.648 &$-$0.513 & 7.052 & 3.888 & 0.722 & 0.684\\
0.04 & 1.664 &$-$0.524 & 7.060 & 3.872 & 0.684 & 0.654\\
0.07 & 1.680 &$-$0.536 & 7.069 & 3.857 & 0.645 & 0.624\\
0.10 & 1.696 &$-$0.548 & 7.077 & 3.842 & 0.607 & 0.594\\
0.13 & 1.712 &$-$0.560 & 7.086 & 3.827 & 0.568 & 0.564\\
0.16 & 1.728 &$-$0.572 & 7.094 & 3.812 & 0.530 & 0.534\\
0.19 & 1.744 &$-$0.583 & 7.103 & 3.797 & 0.491 & 0.504\\
0.22 & 1.760 &$-$0.595 & 7.111 & 3.783 & 0.453 & 0.474\\
\hline
\hline
\end{tabular}
\end{table}

According to B02, the mean extinction correction to RR Lyr, based
on photometric measurements of the astrometric reference stars
located close to this variable, is $<A_V>=0.07\pm0.03$ mag. However,
the same authors also supply an alternative value  $<A_V>=0.11\pm0.10$
mag. Spectroscopic measurements of the diffuse interstellar band at
$\lambda=5780 \AA$ ($E(B-V)=-0.01$ mag) and of the Na D-lines
($E(B-V)=0.12$ mag) carried out by Clementini et al. (1995) do not
allow us to settle the problem of the reddening towards RR Lyr.
On the other hand, empirical relations connecting the intrinsic
color of RR Lyrae
stars to the $B$-band amplitude, the period, and the metallicity
(Piersimoni, Bono, \& Ripepi 2002) or to the Fourier amplitudes of
the $V$-band light curve (Kovacs \& Walker 2001), together with
data listed in Table 5, supply a mean reddening of
$E(B-V)=0.029\pm0.005$ mag. Finally, we note that the color excess
given by Burstein \& Heiles (1978) and by Blanco (1992) is
$E(B-V)=0.03\pm0.02$ mag, while F98 adopts $E(B-V)=0.06\pm0.03$ mag.
As a whole, the weighted mean of extinction corrections available
in the literature gives $<A_V> =0.087\pm0.034$ mag (see arrow
in Fig. 2).

\begin{table}
\centering
\caption[]{Empirical data for RR Lyr based on the
photometry presented by Hardie (1955). The Fourier amplitudes
$A_1$, $A_2$ and $A_3$ were estimated by fitting the $V$-band
light curve by means of a cosine series with 7 components.\label{tab5}}

\begin{tabular}{ll}
P (d)      & 0.5668 \\
(V) (mag)  & 7.784  \\
(B-V) (mag)& 0.371  \\
$A_B$ (mag)& 1.154  \\
$A_V$ (mag)& 0.892 \\
$A_1$      & 0.3014 \\
$A_2$      & 0.1493 \\
$A_3$      & 0.0920 \\
\end{tabular}
\end{table}

By adopting this mean reddening value, the HST astrometric measurement
gives $M_V=0.607\pm0.114$  (intrinsic HST uncertainty) $\pm0.034$ mag
($A_V$ contribution) = $0.607\pm0.119$ mag. On the other hand, the
pulsational approach gives $\log L =1.689 \pm 0.030$, and thus
$M_V=0.624\pm0.041$ (intrinsic uncertainty on the predicted $M_K$
magnitude) $\pm0.044$ mag ($A_V$ contribution)= $0.624\pm0.060$ mag.

The final uncertainty on the predicted RR Lyr parameters appears
still too large to firmly discriminate among current theoretical
or empirical constraints on RR Lyrae intrinsic luminosities and
absolute visual magnitude. However, the use of the $PLZ_K$
relation over a large number of well-studied RR Lyrae stars, see
next section, that cover a wide metallicity range will provide
relevant information on both the slope and, provided that no
systematic errors are affecting the pulsating models, the
zero-point of the $M_V$(RR)-[Fe/H] relation.

\begin{table*}
\caption[]{Observed parameters of RR Lyrae stars for
which accurate visual and near-infrared light curves are available. The
last column lists the sources of the data. The luminosity given in
column (7) was derived using equations (7) [$RRab$] and (8) [$RRc$]
and is in solar units. This luminosity is used in equation (3)
[$RRab$] and (4) [$RRc$] to derive the $M_K$ magnitude. On the basis
of apparent and absolute K magnitudes we evaluate the intrinsic
distance modulus, and in turn the absolute visual magnitude.\label{tab6}}

\begin{tabular}{lccccrrrrc}
\hline
ID  & $\log P$ & [Fe/H] & $E(B-V)$ & $V_0$ & $K_0$ & $\log L$ & $M_K$ & $M_V$ & Ref.$^a$\\

    &          &        &          & (mag) & (mag) &        & (mag) & (mag) & \\

    & & & & & & $\pm$0.04 & $\pm$0.05 & $\pm$0.07 & \\

\hline

\multicolumn{10}{c}{Fundamentals} \\

X Ari & $-0.1863$ & $-2.43$ & 0.15 & 9.078 & 7.894 & 1.852 & $-$0.919 &

0.262 & 2\\

M92 V1 & $-0.1532$ & $-2.24$ & 0.02 & 15.030 & 13.810 & 1.851 & $-$0.965 &

0.255 & 8\\

M92 V3 & $-0.1956$ & $-2.24$ & 0.02 & 15.070 & 13.940 & 1.855 & $-$0.878 &

0.252 & 8\\

SU Dra & $-0.1802$ & $-1.80$ & 0.01 &  9.761 &  8.654 & 1.844 & $-$0.847 &

0.257 & 4\\

VY Ser & $-0.1462$ & $-1.79$ & 0.03 & 10.069 &  8.780 & 1.774 & $-$0.866 &

0.423 & 2\\

RV Oct & $-0.2430$ & $-1.71$ & 0.13 & 10.554 &  9.509 & 1.798 & $-$0.670 &

0.375 & 5\\

RV Phe & $-0.2245$ & $-1.69$ & 0.01 & 11.873 & 10.716 & 1.750 & $-$0.672 &

0.485 & 6\\

RR Leo & $-0.3445$ & $-1.60$ & 0.05 & 10.576 &  9.660 & 1.743 & $-$0.403 &

0.513 & 4\\

W Tuc & $-0.1923$ & $-1.57$ & 0.01 & 11.444 & 10.352 & 1.816 & $-$0.773 &

0.319 & 6\\

TT Lyn & $-0.2237$ & $-1.56$ & 0.01 &  9.833 &  8.630 & 1.711 & $-$0.628 &

0.575 & 4\\

TU UMa & $-0.2536$ & $-1.51$ & 0.02 &  9.764 &  8.656 & 1.728 & $-$0.572 &

0.536& 4\\

SS Leo & $-0.2030$ & $-1.50$ & 0.01 & 11.013 &  9.933 & 1.804 & $-$0.733&

0.347 & 7\\

WY Ant & $-0.2410$ & $-1.48$ & 0.05 & 10.710 &  9.621 & 1.751 & $-$0.612 &

0.477 & 5\\

RR Cet & $-0.2573$ & $-1.45$ & 0.03 &  9.625 &  8.524 & 1.722 & $-$0.552 &

0.549 & 4\\

M5 V8 & $-0.2626$ & $-1.40$ & 0.02 & 15.020 & 13.960 & 1.735 & $-$0.544 &

0.516 & 8\\

M5 V28 & $-0.2645$ & $-1.40$ & 0.02 & 15.040 & 13.990 & 1.739 & $-$0.543 &

0.507 & 8\\

RX Eri & $-0.2312$ & $-1.33$ & 0.05 &  9.529 &  8.358 & 1.699  & $-$0.575 &

0.596 & 4\\

M4 V2 & $-0.2711$ & $-1.30$ & 0.37 & 12.038 & 10.918 & 1.679  & $-$0.473&

0.647 & 9\\

M4 V15 & $-0.3529$ & $-1.30$ & 0.37 & 12.080 & 11.091 & 1.660  & $-$0.287 &

0.702 & 9\\

M4 V32 & $-0.2372$ & $-1.30$ & 0.32 & 11.956 & 10.787 & 1.690  & $-$0.552 &

0.617 & 9\\

M4 V33 & $-0.2112$ & $-1.30$ & 0.30 & 11.906 & 10.745 & 1.726  & $-$0.633 &

0.528 & 9\\

UU Cet & $-0.2175$ & $-1.28$ & 0.03 & 12.008 & 10.841 & 1.713  & $-$0.608 &

0.559 & 6\\

SW Dra & $-0.2444$ & $-1.12$ & 0.03 & 10.389 &  9.326 & 1.728  & $-$0.542 &

0.521 & 2\\

UU Vir & $-0.3228$ & $-0.87$ & 0.01 & 10.534 &  9.516 & 1.637  & $-$0.279 &

0.739 & 4\\

47 Tuc V9 & $-0.1326$ & $-0.71$ & 0.04 & 13.550 & 12.660 & 1.925  & $-$0.871 &

0.019 & 10\\

TW Her & $-0.3984$ & $-0.69$ & 0.05 & 11.091 & 10.217 & 1.615  & $-$0.082 &

0.792 & 2\\

BB Pup & $-0.3180$ & $-0.64$ & 0.10 & 11.838 & 10.902 & 1.669  & $-$0.285 &

0.651 & 5\\

W Crt & $-0.3850$ & $-0.54$ & 0.05 & 11.403 & 10.543 & 1.625  & $-$0.099 &

0.761 & 5\\

DX Del & $-0.3255$ & $-0.39$ & 0.07 &  9.714 &  8.676 & 1.575  & $-$0.168 &

0.870 & 3\\

RS Boo & $-0.4233$ & $-0.36$ & 0.02 & 10.302 &  9.445 & 1.564 & 0.049 &

0.906 & 2\\

AR Per & $-0.3710$ & $-0.30$ & 0.32 &  9.496 &  8.532 & 1.556  & $-$0.048 &

0.916 & 4\\

RR Gem & $-0.4009$ & $-0.29$ & 0.08 & 11.113 & 10.208 & 1.555 & 0.017 &

0.922 & 4\\

SW And & $-0.3543$ & $-0.24$ & 0.06 &  9.510 &  8.508 & 1.547  & $-$0.069 &

0.933 & 3\\

V445 Oph & $-0.4010$ & $-0.19$ & 0.27 & 10.131 &  9.151 & 1.500 & 0.070 &

1.050 & 7\\

AV Peg & $-0.4085$ & $-0.08$ & 0.07 & 10.274 &  9.309 & 1.489 & 0.108 &

1.073 & 4\\

RR Lyr & $-$0.2466 &$-$1.39 & 0.07 & 7.714 & 6.532 & 1.680 & $-$0.536 &

0.645 & $^b$\\

\multicolumn{10}{c}{First Overtones} \\

TV Boo & $-0.5051$ & $ -2.44 $ & 0.00 & 10.984 & 10.200 & 1.770 & $-$0.393 &

0.391 & 4\\

T Sex & $-0.4885$ & $-1.34$ & 0.05 &  9.886 &  9.156 & 1.729 & $-$0.299 &

0.431 & 4\\

DH Peg & $-0.5926$ & $-1.24$ & 0.08 &  9.287 &  8.587 & 1.613 & 0.019 &

0.719 & 1\\

\hline

\hline

\end{tabular}

$^a$ References: 1) Jones et al. 1988a; 2) Jones et al. 1988b; 3) Jones

et al. 1992; 4) Liu \& Janes 1990a; 5) Skillen et al. 1993; 6)

Cacciari et al. 1992; 7) Fernley et al. 1990; 8) Storm et al.

1994a; 9) Liu \& Janes 1990b; 10) Storm et al. 1994b.

$^b$ See text for more details.

\end{table*}

\section{Field RR Lyrae stars}

Following the procedure discussed in the previous section for RR
Lyr itself, the intrinsic luminosity of individual RR Lyrae stars
with well-determined $K$ and $V$ magnitudes, as well as
metallicity and reddening, can be directly inferred from the
measured $V-K$ color, without any {\em a priori} assumption
related with the predicted luminosity of horizontal branch models. 
Once the individual luminosity of $RRab$ variables is derived from
equation (7) (equation (8) for $RRc$ variables, the predicted
$PLZ_K$ relations [see equations (3) and (4)] provide the absolute
near-infrared magnitude. This supplies the intrinsic distance
modulus, and in turn the absolute visual magnitude. This means
that we can investigate both the $M_K$-[Fe/H]-log$P$ and the
$M_V$-[FeH] relations.

We searched for RR Lyrae stars for which accurate $V$ and $K$
light curves, as well as reliable evaluations of both reddening
and metallicity, are available in the literature. A total of 39
stars were found, with 30 field and 9 GGC members. Metal
abundances have been taken from the homogeneous compilation by
Fernley et al. (1998a). The sample is listed in Table 6 together
with key parameters such as the luminosity estimated according to
equation (7) or (8), the $M_K$ provided by equation (3) or (4),
and the ensuing $M_V$ magnitudes. The errors associated with
luminosity, near-infrared, and visual absolute magnitudes account
for the intrinsic dispersion of the predicted relations, as well
as for an average uncertainty of $\pm0.20$ dex on the metallicity
and of $\pm0.03$ mag on $A_V$.

Figure 3 shows the derived "pulsational" luminosity of RR Lyrae
stars in our sample as a function of the metal content. Filled and
open circles display field $RRab$ and $RRc$, respectively. $RRab$
stars in GGCs (M4, M5, and M92) are plotted as filled triangles,
while RR Lyr itself with an asterisk. The open triangle marks the
position of the RR Lyrae V9 in 47 Tuc which was neglected, since
it is peculiar (Storm et al. 1994b). Together with individual RR
Lyrae stars, Fig. 3 also shows the predicted $\log L_{RR}-$[Fe/H]
relation according to the old ZAHB models by Castellani, Chieffi,
\& Pulone (1989, 1991, hereinafter CCP, dotted line) as well as
the improved computations presented by Cassisi et al. (1998, 1999,
hereinafter C99, solid line). The dashed line displays the locus
of central He exhaustion according to C99.

\begin{figure}
\psfig{figure=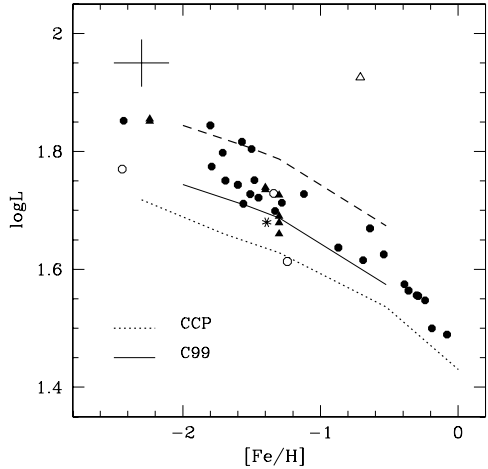,height=9cm}
\caption{Estimated individual luminosities of RR Lyrae stars in our sample
versus metal abundance (see Table 6). Filled and open circles show
field $RRab$ and $RRc$ variables, while filled triangles indicate cluster
$RRab$ variables. The asterisk and the open triangle show RR Lyr
itself and the peculiar RR Lyrae V49 in 47 Tuc, respectively.
The error bars display current uncertainties in metal abundance and in 
luminosity.  Dotted and solid lines refer to old and recent ZAHB models, 
respectively, while the dashed line depicts the exhaustion of the central 
He-burning phase. See text for more details.}
\end{figure}

As a whole, pulsational luminosities appear in reasonable agreement 
with the result of recent evolutionary computations based on improved 
input physics. In fact, with the exception of V9 in 47 Tuc, they show 
a dispersion in luminosity of the order of 0.10 dex (dashed line) above 
the theoretical ZAHB, in agreement with the predictions of synthetic 
horizontal branch simulations (see.  e.g. Caputo et al. 1993; 
Demarque et al. 2000). Thus it seems that recent evolutionary 
computations and pulsational constraints are supporting each other.

When passing to magnitudes, one has to recall that stellar
evolution predicts the luminosity and, therefore, the magnitude of
"static" stellar structures, whereas the present pulsational
scenario refers to magnitude-weighted mean magnitudes. As
discussed by Bono, Caputo \& Stellingwerf (1995), variables that
present large luminosity amplitudes are expected to have mean
visual magnitudes up to $\sim$ 0.15 mag fainter than the
equivalent static values. Bearing in mind this {\em caveat}, Fig.
4 shows the predicted (magnitude-weighted) absolute visual
magnitudes of the variables in our sample as a function of
metallicity. The comparison with the ZAHB theoretical predictions
by Cassisi et al. (1998, 1999) is now less meaningful and only
discloses the compatibility of pulsational and evolutionary
scenarios, perhaps  suggesting that the evolutionary models are
somehow too bright with respect to pulsational magnitudes.
However, a more detailed investigation concerning the comparison
with current evolutionary predictions requires reliable amplitude
corrections to the observed mean magnitudes to obtain the
equivalent static values.

\begin{figure}
\psfig{figure=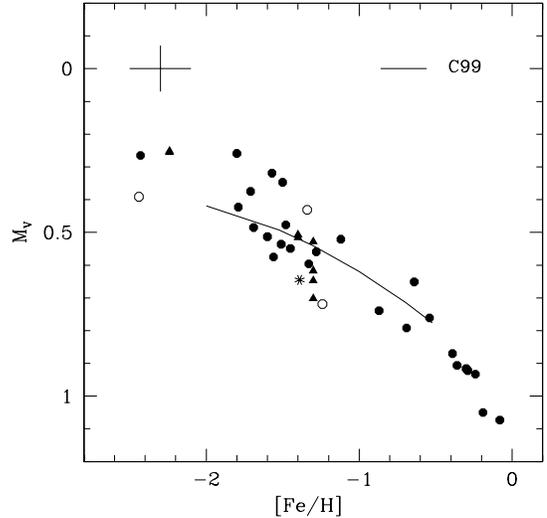,height=10cm}
\caption{Pulsational absolute visual
magnitudes of RR Lyrae variables in our sample as a function of metal
content. The symbols are the same as in Fig. 3.}
\end{figure}

It is worth noting that a least-squares solution through the data
plotted in Fig. 4 suggests that the $M_V$-[Fe/H] relation becomes
steeper and steeper when moving from metal-poor to metal-rich
variables (see Table 7). This result is in agreement with the
change in the slope at [Fe/H]$\sim -$1.5 originally predicted by
CCP and further supported by synthetic HB simulations (Caputo et
al. 1993; Demarque et al. 2000), by RR Lyrae stars in the field 
(MacNamara 1999), in GGCs (Caputo et al. 2000), and in $\omega$
Centauri (Rey et al. 2000).

\begin{table}
\center \caption[]{The slope of the $M_V$(RR)-[Fe/H] relation as a
function of the adopted metallicity range.\label{tab6}}
\begin{tabular}{cc}
\hline
[Fe/H] range & slope\\
\hline
$-$2.4/$-$1.8 & $\approx$ 0\\
$-$2.4/$-$1.7 & 0.080$\pm$0.100\\
$-$2.4/$-$1.6 & 0.190$\pm$0.090\\
$-$2.4/$-$1.5 & 0.192$\pm$0.079\\
$-$2.4/$-$1.4 & 0.221$\pm$0.060\\
$-$2.4/$-$1.3 & 0.282$\pm$0.053\\
$-$2.4/$-$1.0 & 0.284$\pm$0.050\\
$-$2.4/$-$0.6 & 0.283$\pm$0.039\\
$-$2.4/$-$0.3 & 0.303$\pm$0.022\\
\hline
\hline
\end{tabular}
\end{table}

\section{$PLZ_K$ versus FOBE and BW absolute magnitudes}

Figure 5 shows the comparison between the absolute visual
magnitudes of field RR Lyrae stars estimated using the predicted
$PLZ_K$ relation (circles, see column (9) in Table 6) and the
averaged absolute visual magnitudes of RR Lyrae stars in GGCs
(triangles) derived by Caputo et al. (2000) according to the
"First Overtone Blue Edge" (FOBE) method. The agreement between
these two independent approaches is excellent, thus supporting the
consistency of the pulsational scenario. At the same time,this
result indicates that field and cluster RR Lyrae stars should have
experienced  quite similar evolutionary histories.

As a whole, linear interpolation through the data plotted in Fig. 5
gives
the following analytical relations (solid lines):

$$M_V=0.718(\pm0.072)+0.177(\pm0.069)\FeH\eqno(9)$$

\noindent
for RR Lyrae stars with [Fe/H]$\le-$1.6, and

$$M_V=1.038(\pm0.077)+0.359(\pm0.027)\FeH\eqno(10)$$

\noindent
for RR Lyrae stars with [Fe/H]$>-$1.6.

\begin{figure}
\psfig{figure=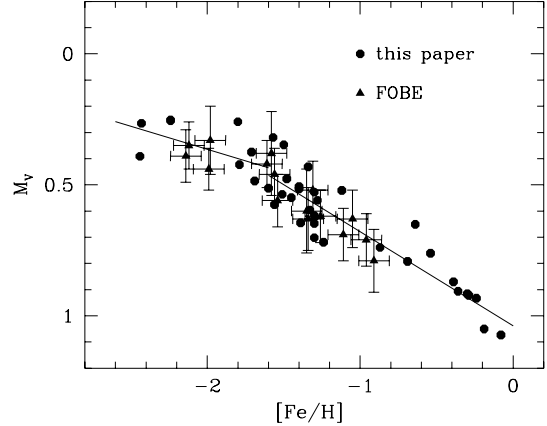,height=10cm}
\caption{Comparison between absolute visual magnitudes of RR Lyrae
variables in our sample evaluated using the current pulsation approach
(circles) and the results of First Overtone Blue Edge (FOBE) method (triangles)
to globular cluster variables (Caputo et al. 2000). The solid lines are the
relations described in the text.}
\end{figure}

Let us now compare our results with current distance
determinations based on the BW method. To perform a robust
comparison we adopted the absolute visual magnitudes for field RR
Lyrae stars derived by Fernley (1994). The key feature of this
sample is that the empirical data available in the literature were
reduced by the previous author to the same NIR BW method. At the
same time, for these objects the metal abundances were also
estimated by adopting a homogeneous metallicity scale (Fernley et
al. 1998a). We ended up with a sample of 27 RR Lyrae stars. The
nine GGC variables were not included, since they appear to be
affected by larger intrinsic errors and also because they have not
been reduced to the same BW method yet. Figure 6 shows the
comparison between these NIR BW absolute magnitudes (filled
triangles) and the absolute visual magnitudes for the same
objects, as based on the $PLZ_K$ relation (circles). Data plotted
in this figure display quite clearly that for RR Lyrae stars more
metal-poor than [Fe/H] $\sim -1.5$, the discrepancy is on average
of the order of 0.2 mag, in the sense that BW magnitudes are
fainter than those provided by current pulsational approach. The
discrepancy decreases toward higher metal contents and vanishes in
the metal-rich tail. This indicates that the metallicity
dependence given by the BW method is milder than predicted by
pulsation models.

\begin{figure}
\psfig{figure=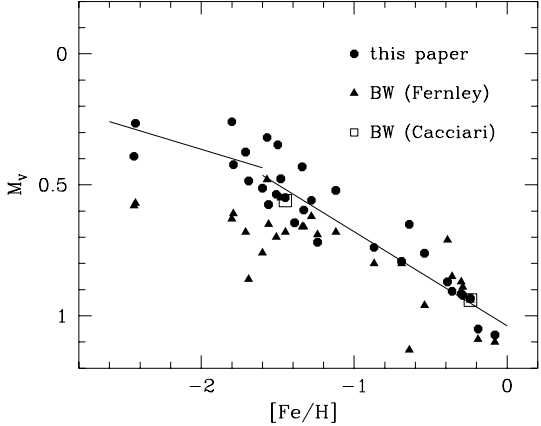,height=10cm}
\caption{Comparison between the absolute visual magnitudes estimated
using the current pulsation approach and the independent evaluations
derived by Fernley (1994) using the NIR BW
method (triangles). The two open squares mark the position of
RR Cet ([Fe/H]=$-1.45$) and SW And ([Fe/H]=$-0.24$) according to the
new calibration of the BW method provided by Cacciari et al. (2000).
The solid lines are the
relations described in the text.}
\end{figure}

Note that a new calibration of the BW method has been provided by
Fernley et al. (1998b), but the difference over the entire
metallicity range is negligible when compared with Fernley (1994).
On the contrary, the preliminary analysis provided by Cacciari et
al. (2000) of the assumptions currently adopted in the BW method
disclosed quite different results. In particular, by using
available photometric data together with a large set of new
atmosphere models, they found that RR Cet ([Fe/H]=$-1.45$) is
$\approx$ 0.12 mag brighter than previously estimated, whereas no
significant correction was required for SW And ([Fe/H]=$-0.24$).
As a result, these new absolute magnitudes (open squares in Fig.
6), once confirmed, appear fully consistent with the $M_V-$[Fe/H]
relation based on pulsation models [equation (10)]. This suggests
that the slope of this relation, at least in the metallicity range
$-1.5\le$[Fe/H]$\le -0.4$, is steeper than currently adopted (see
also McNamara 1999).

Following a referee's suggestion, one may investigate whether 
the discrepancy  at [Fe/H] $< -$ 1.5 between pulsational absolute
visual magnitudes and BW magnitudes might be settled by changing
the mass value in equations (1) and (5). We found that to remove
the difference of $\sim$ 0.25 mag between BW and pulsational
magnitudes at [Fe/H]=$-$1.8, one should account for a decrease 
in mass of the order of $\sim$ 23\%, thus predicting a quite 
unrealistic low-mass, 0.55$M_{\odot}$, for these metal-poor 
($Z\sim$ 0.0003) horizontal branch stellar structures.

As for $K$-magnitudes, Fig. 7 shows that the absolute $K$-magnitudes 
estimated using the pulsational approach do obey a very tight relation 
with the metal content, once the dependence on period is removed. 
Owing to the small number (3) of $RRc$ variables, their periods have 
been fundamentalized, (log$P_F$=log$P_{FO}$+0.127), thus yielding for 
the entire sample the following relation:

$$M^F_{K}=-0.770(\pm0.044)-2.101\log P+0.231(\pm 0.012)\FeH\eqno(11)$$

The small errors affecting the coefficients of this relation further

strengthen the evidence that near-infrared observations could supply
accurate distance determinations.

\begin{figure}
\psfig{figure=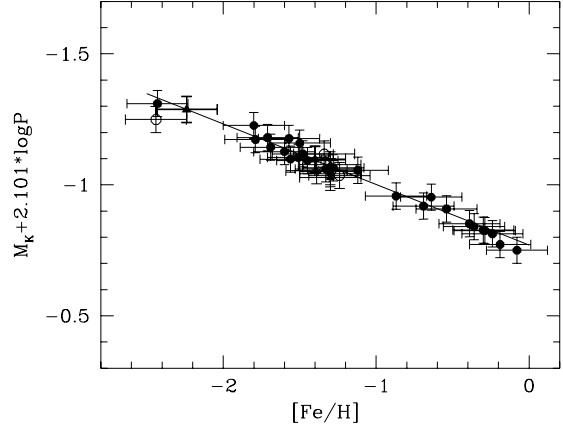,height=10cm}
\caption{Absolute $K$-magnitudes of RR Lyrae in our sample estimated
using the pulsation method as a function of metal content. Data were
projected onto a two dimensional plane. The solid line shows the
projection onto the same plane of equation (11). The symbols are
the same as in Fig. 3.}
\end{figure}

To investigate in more detail the discrepancy between the current
approach and the BW method, we compare the pulsational absolute
$K$ magnitudes with the absolute $K$ magnitude obtained using the
BW true distance modulus and the dereddened apparent K magnitudes
listed in column (6) of Table 6. Figure 8 shows the comparison
between pulsational $M_K$ values (circles) and results based on
the BW method (triangles). Once again, the BW magnitudes at
[Fe/H]$\le-$ 1.5 are fainter than predicted by the pulsational
approach, leading to a milder dependence on [Fe/H] with respect to
the $PLZ_K$ predictions (solid line). However, the absolute
$K$-magnitudes derived by Cacciari et al. (2000) for SW And and RR
Cet (open squares) appear once again in good agreement with our
pulsational approach.

\begin{figure}
\psfig{figure=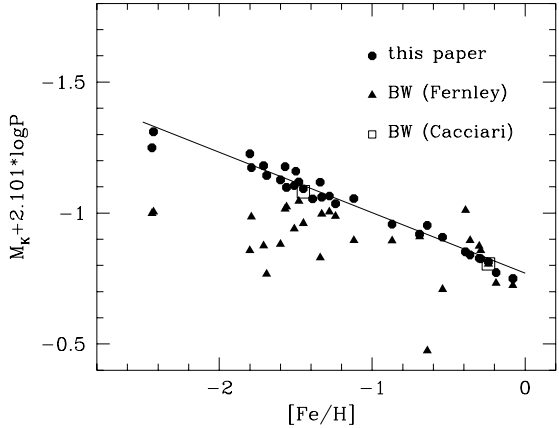,height=10cm}
\caption{Comparison between the absolute
$K$-magnitudes based on the pulsation approach and the BW method.
The latter were derived using the true distance moduli
given by Fernley (1994) and the $K_0$-magnitudes listed
in Table 6. The solid line refers to equation (11).}
\end{figure}

\section{Summary and final remarks}

Data plotted in Fig. 6 and Fig. 8 clearly show that the pulsational
approach provides absolute magnitudes for RR Lyrae stars that are
significantly brighter than suggested by the BW method. This difference 
is more evident for metal-poor ([Fe/H]$\le-$ 1.5) RR Lyrae variables.
The recent revision provided by Cacciari et al. (2000) for RR Cet 
([Fe/H]=$-$1.45) and SW And ([Fe/H]=$-$0.24) agrees quite well with the 
pulsational results. This suggests that both the zero-point and the slope 
of the $M_V$-[Fe/H] and of the $M_K$-[Fe/H]-log$P$ relation based on the 
old BW calibration need to be revisited.

We have also found that both $V$ and $K$ absolute magnitudes of RR Lyr 
itself, estimated via the pulsational approach, are in good agreement 
with the trigonometric parallax recently measured by HST (see Fig. 2). 
This suggests that at least for the metallicity of RR Lyr ([Fe/H]=$-$1.39), 
the pulsational approach is consistent with direct distance determinations.

\begin{figure}
\psfig{figure=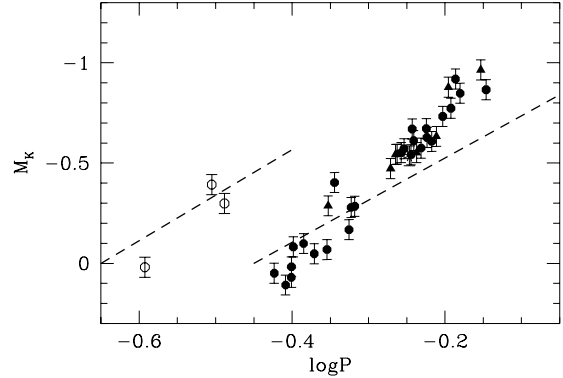,height=10cm}
\caption{Absolute $K$-magnitudes for the current sample of well-observed
field and cluster RR Lyrae stars. Symbols are the same as in Fig. 3.
The dashed lines depict the predicted slope at constant luminosity and 
metal content.}
\end{figure}

As far as the metal-dependence is concerned, we finally note that
the effect of metallicity on equation (11) can be revealed by a
simple cross-check between predicted $K$ magnitudes and periods.
It has been already shown (see Paper I) that $K$ measurements for
RR Lyrae stars in M3, namely for variables at quite constant
luminosity and metal content, closely follow the pulsational
constraints on the period dependence. On the contrary, data
plotted in Fig. 9 show that for the current sample of cluster and
field RR Lyrae stars the $M_K$-log$P$ relation is significantly
steeper than predicted under the above quoted assumptions (dashed
lines). This is the expected aftermath of the fact that metal-poor
variables, which have longer periods (see data in Table 6), have
also brighter bolometric luminosities (see Fig. 3) than metal-rich
ones.

{\bf Acknowledgments}: It is a pleasure to thank C. Cacciari and
J. Lub for many useful discussions on the Baade-Wesselink method
and on current reddening corrections. We wish to warmly thank the
second anonymous referee for several comments and suggestions that
improved the content and the readability of the paper. Financial
support for this work was provided by MIUR-Cofin 2000, under the
scientific project "Stellar Observables of Cosmological Relevance"
(V. Castellani and A. Tornamb\'e, P.I.)

\end{document}